\documentclass[twocolumn,showpacs,preprintnumbers,amsmath,amssymb]{revtex4}

\usepackage{graphicx}
\usepackage{dcolumn}
\usepackage{bm}

\begin{document}

\preprint{APS/123-QED}

\title{Actively-Induced Percolation: An Effective Approach to \\
Multiple-Object Systems Characterization}

\author{Luciano da Fontoura Costa}
\affiliation{
Institute of Physics of S\~ao Carlos. 
University of S\~ ao Paulo, S\~{a}o Carlos,
SP, PO Box 369, 13560-970,
phone +55 162 73 9858,FAX +55 162 71
3616, Brazil, luciano@if.sc.usp.br
}%

\date{10th April 2004}

\begin{abstract}

The present work proposes the concept of induced percolation over
multiple-object systems, so that features such as the number of merged
clusters can be used as a relevant measurement.  The suggested
approach involves the expansion of the objects while monitoring the
evolving clusters.  The potential of the proposed methodology for
characterizing the spatial interaction and distribution between
several objects is illustrated with respect to synthetic and real
data.

\end{abstract}



\maketitle

Great part of the current interest in statistical physics and complex
systems targets critical phenomena, such as percolation
\cite{Stauffer:94}.  Characterized by abrupt changes of the system
properties as some parameter is varied, critical phenomena are of
interest because of their rich and complex dynamics.  As a matter of
fact, while the calculation of critical exponents for continuous phase
transitions has been identified as one of the greatest achievements of
theoretical physics over the last 25 years \cite{Binney_etal:94}, the
study of geometric percolation in graphs has provided one of the main
motivations of the new area of complex networks
\cite{Albert_barab:2002}.  Much interest has been focused on the
analysis, modeling and simulation of natural critical phenomena,
whereas relatively little attention has been given to
\emph{artificially-induced critical phenomena}, namely the enforcement
of the objects in the system to undergo some imposed dynamics ---
e.g. expansion or dilation --- in terms of a control parameter, so
that critical phenomena can be induced.  Although the intrinsic
relevance of such investigations is potentially limited by the imposed
evolution dynamics, it is argued and illustrated in the current work
that such experiments present great potential for the characterization
of the spatial distribution of objects while taking into account their
shape and topology.  This perspective provides a novel approach to the
old problem of geometric characterization of multiple-object physical
systems, which has been approached from several perspectives including
densities (e.g. \cite{Wassle:78, Ziman:79}), additive functionals
(e.g. \cite{Raedt01}), fractal dimension
(e.g. \cite{Kaye:94,Stoyan:95}) and lacunarity (e.g. \cite{Hovi:96}).
The importance of obtaining an effective methodology for the
characterization of the spatial distribution and interaction in such
systems is greatly enhanced by the myriad of related physical and
biological systems, including statistical physics, astrophysics,
geology, material science and even genetics (e.g. the spatial patterns
of gene expression \cite{Streicher:2000}).  Although the proposed
methodology can be immediately extended to continuous spaces, the
following developments are limited to discrete structures as these are
naturally required by several computational implementations and
applications. In this article we start by presenting the proposed
induced-percolation concepts and methods and proceed by validating and
illustrating them with respect to uniform point densities, perturbed
hexagonal lattices and real data regarding the spatial distribution of
retinal ganglion cells (e.g. \cite{Wassle:78}).

\begin{figure}
 \begin{center} 
   \includegraphics[scale=.5,angle=-90]{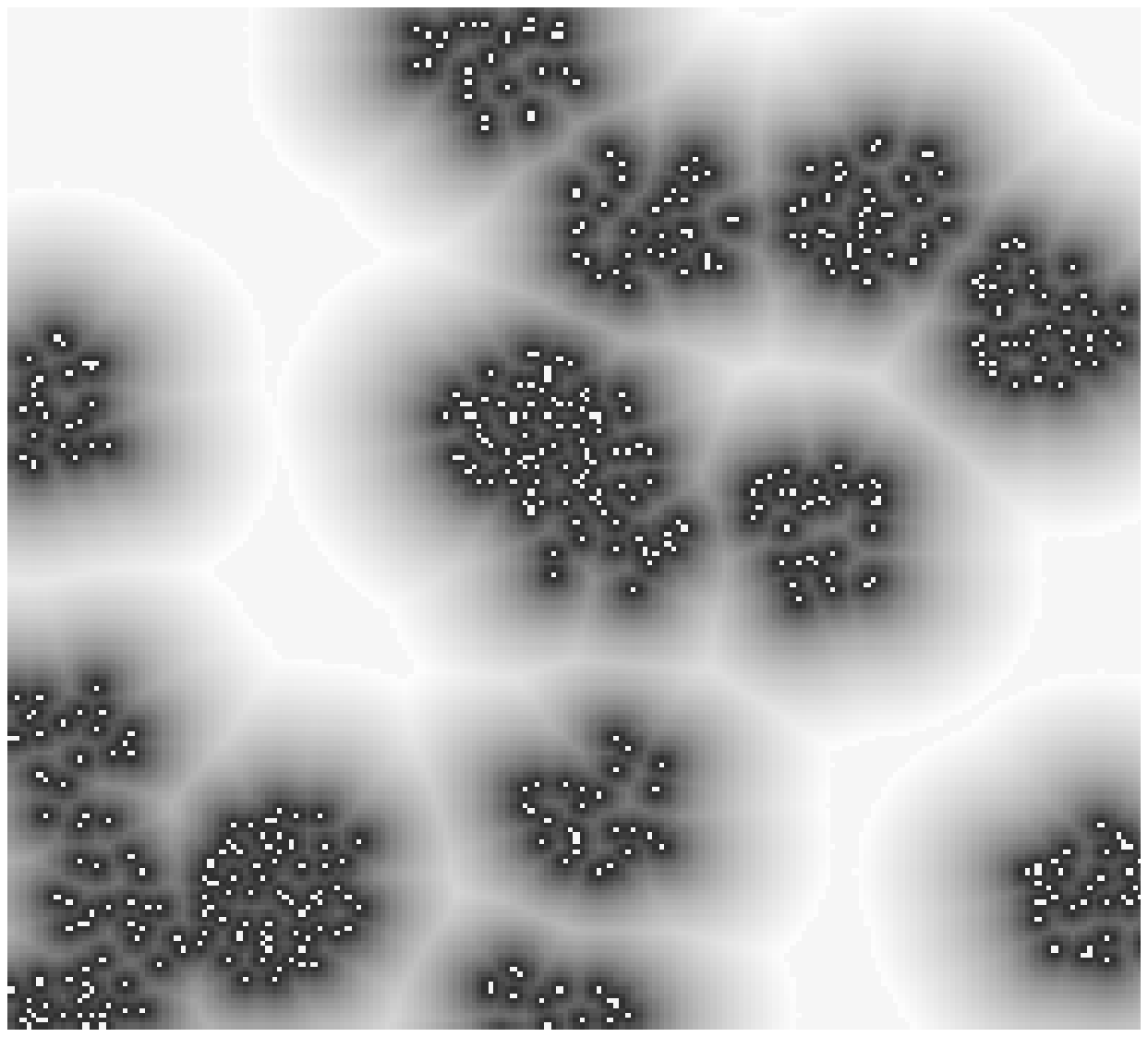} \\
   \vspace{0.2cm} (a) \\
   \includegraphics[scale=.55,angle=-90]{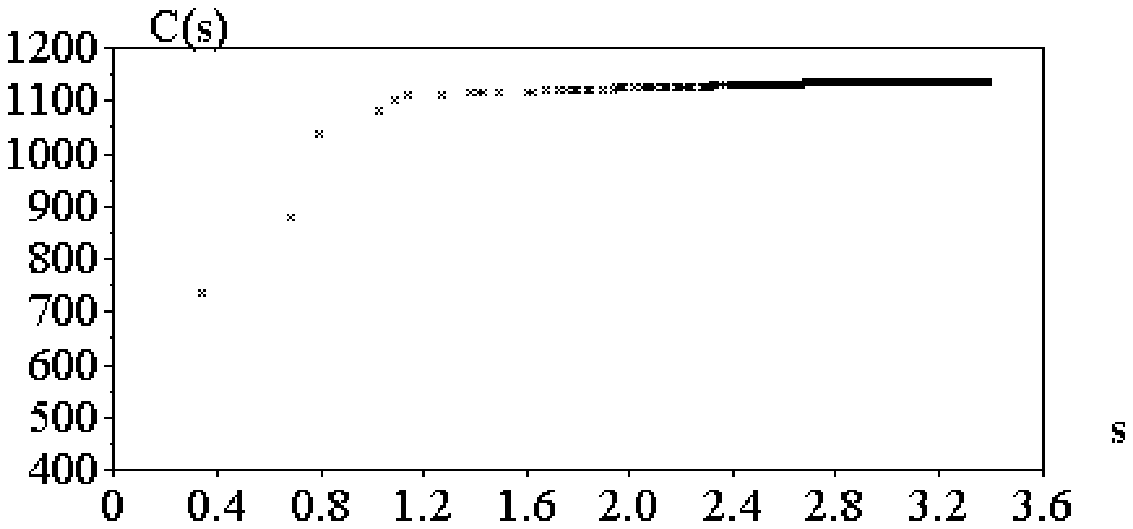} \\
   (b) \\
   \includegraphics[scale=.55,angle=-90]{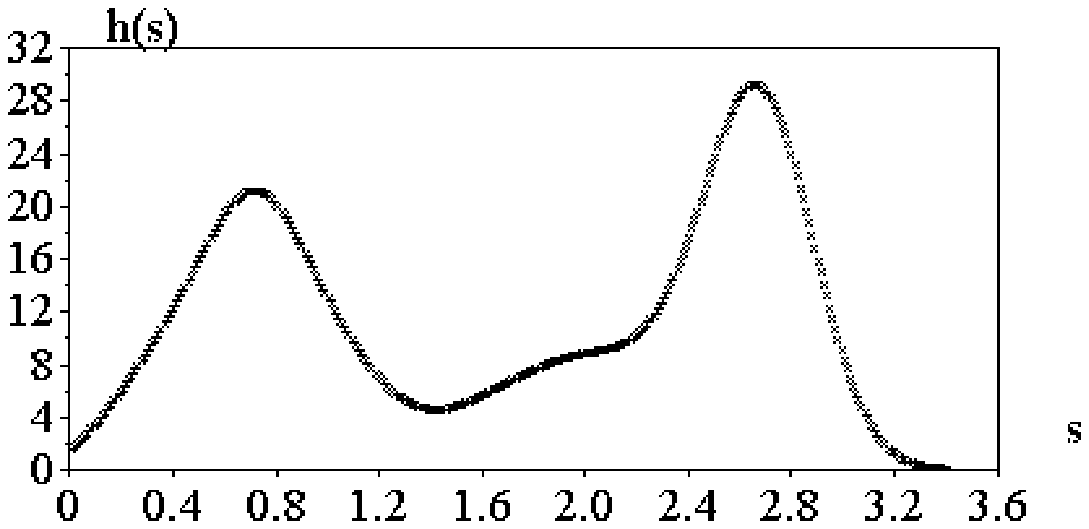} \\
   (c)

   \caption{The dilations of co-existing uniform particle densities
   (a) and the respective number of mergings $C(s)$ and $h(s)$
   function. ~\label{fig:ex}} \end{center}
\end{figure}

Let $\Omega$ be the $N-$dimensional Euclidean space where the
multiple-object system is embedded and $S$ be the set containing the
coordinates $(x,y)$ of each point of each object, in any order.  The
definition of the subsets of points corresponding to each object can
be done by considering the respective point connectivity or by
imposing previous knowledge about their classes (in this case a single
object may include more than one connected component).  In this work,
every pair of points $a$ and $b$ such that $|x_a-x_b|+|y_a-y_b|=1$ is
linked through an undirected edge, so that each object corresponds to
a connected component (or cluster) in $\Omega$.  Let also $R \in
\Omega$ be a set delimitating the region of $\Omega$ where the
artificially-induced dynamics is allowed.  The Euclidean $D$-ring of
the objects in $\Omega$ is obtained by determining the set of points
of $R$ which are no further away than $D$ from the set $S$
\footnote{Recall that the distance between a point $p$ and a set $S$
of points corresponds to the minimal distance between $p$ and any of
the elements of $S$.}.  Only the distance values $d_i$ allowed in the
orthogonal lattice, i.e. $d_i=\sqrt{x^2+y^2}$ for some $x,y$ belonging
to $\Omega$, are considered in order to obtain full precision in
spatially discrete spaces \cite{CostaCesar:2001}.  Observe that the
density of such possible distances tends to increase with the
distance.  The dilation of $S$ by the distance $d_i$ can now be
defined as the set of points obtained by the union of the original
object and the respective $D=d_i$-ring.  Figure~\ref{fig:ex} shows a
multiple-object system (a), the dilation of such a system for
distances up to $D=10$ (b), which can also be understood in terms of
the exact distance transform of the objects \cite{CostaCesar:2001}.

The percolation dynamics considered in the present work involves the
dilation, up to a maximum distance $d_M$, of the objects by each
subsequent exact distance $d_i$, a scheme henceforth called
\emph{exact dilations} \cite{CostaCesar:2001}.  The exact distance
$d_i$, or its respective index $i$
\footnote{Observe that this index corresponds to the order of the
exact distances $d_i$ sorted in ascending order.  For instance, in the
case of the orthogonal lattices, we have $d_0=0$, $d_1=1$,
$d_2=\sqrt{2}$, etc.}, can both be considered as percolation
parameter.  Two are the immediate consequences of such an enforced
dilation procedure: (i) the area of the objects increases
monotonically; and (ii) contacts between objects occur at specific
distance values, inducing the respective merging of the involved
clusters and possible critical phenomena.  By associating each object
to a node and incorporating edges between pairs of objects whenever
they first touch one another, it is possible not only to merge the
objects into larger clusters, but also to obtain the hierarchy induced
by this agglomerative procedure.

Several measurements, such as the size of the maximum cluster
\cite{Stauffer:94, Albert_barab:2002}, could be considered in order to
identify possible critical phenomena while the parameter $i$ is
increased, but the current work focuses on the cumulative number of
clusters merged at each successive distances $d_i$, which is
henceforth expressed as $C(d_i)$.  We consider a square window with
area $A$. In order to provide spatial-scale uniformity, we make the
change of variable $C(k_i=log(d_i))$.  Now, considering uniform
distribution of points (i.e. Poisson) with density $\gamma$ as a
reference model, the rate of change at $k_j$, while going from
$C(k_j)$ to $C(k_{j+1})$, can be maximized by the finite difference
$q(k_j) = P(k_j)/ \Delta k_{j}$, where $P(k_j)=\gamma A$ is the
expected number of points inside $R$ and $\Delta k_{j} = k_{j+1}-k_j$.
In order to have the same change $q(k_i)$ for any value of $k_i$ while
considering uniform point densities inside the fixed window, we choose
the changing rate at any specific value of $k_j$ (in this article
$j=1$) as a reference and make the normalization as given in
Equation~\ref{eq:norm}.  By adopting the minimal nearest neighbor
distance as the prototype distance for the uniform distribution with
density $\gamma$, we can use the mean expected value of such a
distance, given as $d_i = exp(k_i) = 1/(2 \sqrt{\gamma})$
\cite{Stoyan:95}, in order to relate $\gamma$ and $k_i$.  By using
this expression, we have that $P(k_i)=\gamma A = A/(4d_i^2)$ and $
P(k_j) = A/(4{d_j}^2)$, so that $P(k_j)/P(k_i) = exp(2k_i-2k_j)$ and
it is now possible to rewrite Equation~\ref{eq:norm} as
Equation~\ref{eq:norm1}.  Now, the derivative of the number of cluster
mergings in terms of $k_i$ for any type of point distribution can be
approximated as in Equation~\ref{eq:w}, which is used henceforth.
This equation already takes into account the fact that $k_{j=1}=0$. An
interpolated version of $w(k_i)$ can be obtained by first embedding it
onto the continuous space $s$, such that $s=k_i$ at $i$, as expressed
in Equation~\ref{eq:ws}, and subsequently using the Parzen
interpolating scheme \cite{DudaHart:2001} given in
Equation~\ref{eq:parzen} to obtain the function $h(s)$, where
$g_{\sigma}(s)=1/ \sqrt{2 \pi}/ \sigma exp(-0.5(s/ \sigma)^2)$ is the
normal density with standard deviation $\sigma$ and `*' stands for
convolution.  Figures~\ref{fig:ex}(b) and (c) illustrates the
distributions $C(s)$ and $h(s)$, respectively, for the multiple-object
system in (a).  The two co-existing Poisson densities in
Figure~\ref{fig:ex}(a), which were generated with $\gamma_1=0.05$ and
$\gamma_2=0.002$, are clearly identified by the two peaks obtained in
function $h(s)$.  Each of these peaks have respective maximum value at
$s_1=0.728$ and $s_2=2.673$, yielding estimated densities
$\hat{\gamma_1}=1/(4*exp(2s_1))=0.058$ and $\hat{\gamma_2}=0.0012$.
Considering the interferences between the two co-existing points
distribution and the relatively poor sampling of the less dense
distribution by the adopted $400 \times 400$ points window, such
results can be considered as being reasonably accurate.  Moreover, the
fact that only a fraction of the denser distribution present in the
window is duly expressed by the lower peak obtained for that case.
Observe that such a multimodal distribution could by no means be
obtained by using nearest-neighbor distance statistics.

\begin{eqnarray}
  Q(k_i)= \frac{P(k_i)}{\Delta k_i} \left( \frac{\Delta k_i}{\Delta k_j} \frac{P(k_j)}{P(k_i)} \right)  \label{eq:norm}  \\ 
  Q(k_i)= \frac{P(k_i)}{\Delta k_j} exp(2k_i-2k_j)  \label{eq:norm1}  \\ 
  w(k_i)= \left[ C(k_{i+1})-C(k_i)) \right] / \Delta k_1 exp(2k_i)  \label{eq:norm2}  \label{eq:w} \\
  w(s) = \sum_{i=1}^{M} w(k_i) \delta(s-d_i) \label{eq:ws} \\
  h(s) = \sum_{i=1}^{M} \left[ w(k_i) \delta(d_i) \right]* g_\sigma(s)  \label{eq:parzen} \\
  d(s) = 2 \gamma \pi exp(2s- \gamma \pi e^{2s})  \label{eq:ds}
\end{eqnarray}

\begin{figure*}[t]
 \begin{center} 
   \includegraphics[scale=.6,angle=-90]{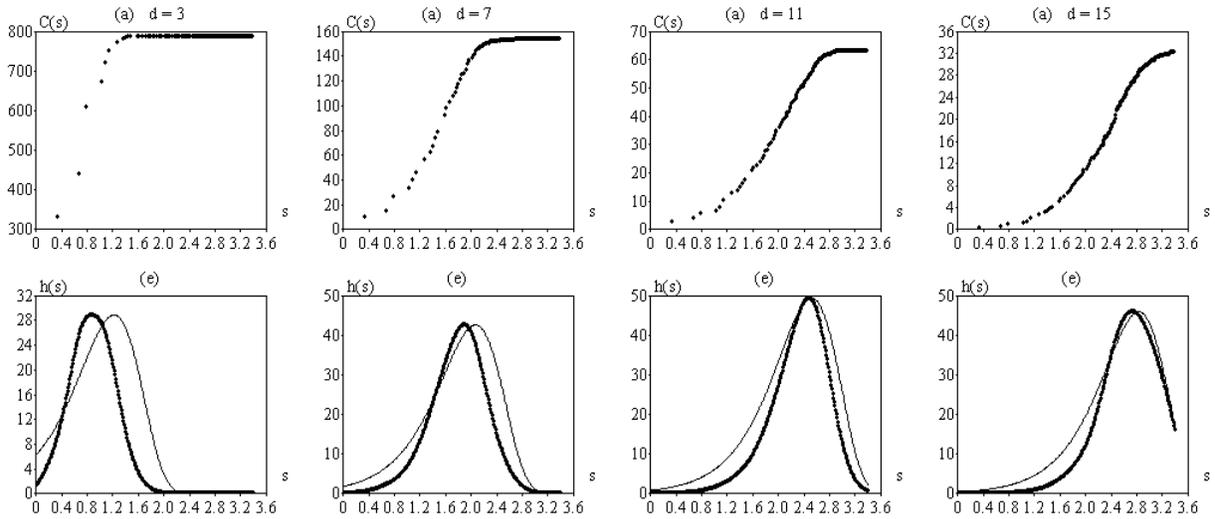} \\

   \caption{Number of cluster mergings $C(s)$ (a-d) and respective
   $h(s)$ functions (e-h), represented by thick lines, obtained for
   uniform point densities with typical spatial scales 3, 7 11 and
   15. The respective nearest-neighbor distributions are shown by the
   solid curves~\label{fig:dens}} \end{center}
\end{figure*}

In order to further validate the proposed methodology, it has been
applied to a series of uniform point distributions with increasing
densities $\gamma = 1/(4d^2)$, where $d=1, 2, \ldots, 15$ are the
characteristic spatial scales, estimated by nearest-neighbor
distances.  A $200 \times 200$ points window was adopted.  The cases
obtained for $d=3, 7, 11$ and 15 are shown in Figure~\ref{fig:dens}.
Two interesting results have been identified: (i) a reasonably good
estimation of the typical densities can be obtained by using the
suggested methodology; and (ii) the functions $h(s)$ practically do
not change their shape or heights for the different cases.  It is also
interesting to observe that the obtained functions are similar, but
not identical, to the nearest-neighbor distributions given by
Equation~\ref{eq:ds}, also presented in Figure~\ref{fig:dens}.

The potential of the proposed methodology for quantifying the degree
of disorder \cite{Ziman:79} in point distributions was also evaluated
by considering hexagonal lattices with increasing levels of uniformly
distributed noise.  More specifically, the basic hexagonal
distribution was characterized by basic cells with sides of 10 points,
with progressive perturbations added to the points coordinates. A $200
\times 200$ window was also considered for this case.  It is clear
from the obtained results that the perturbation degree is immediately
reflected by the dispersion of the obtained functions $h(s)$, while
the typical point density (indicated by the position of the peaks) is
kept fixed.  The sharpest peak is obtained for the noiseless hexagonal
structure.  The nearest-neighbor distributions for respective density
is also shown (solid line) superposed to the $h(s)$, whose relatively
narrow dispersion indicates that the point system is far from the
disorder characteristic of uniform distributions.

\begin{figure}
 \begin{center} \includegraphics[scale=.6,angle=-90]{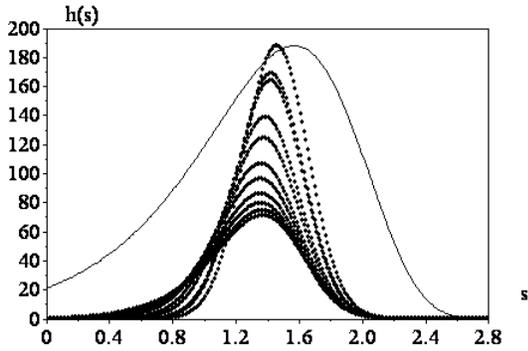} \\

   \caption{Profiles of $h(s)$ (dotted lines) obtained for hexagonal
   point distributions with increasing levels of perturbation. The
   nearest-neighbor distribution obtained for uniform point dispersion
   with equivalent density is shown by the solid curve.
   ~\label{fig:hex}} \end{center}
\end{figure}

\begin{figure}
 \begin{center} 
   \includegraphics[scale=.6,angle=-90]{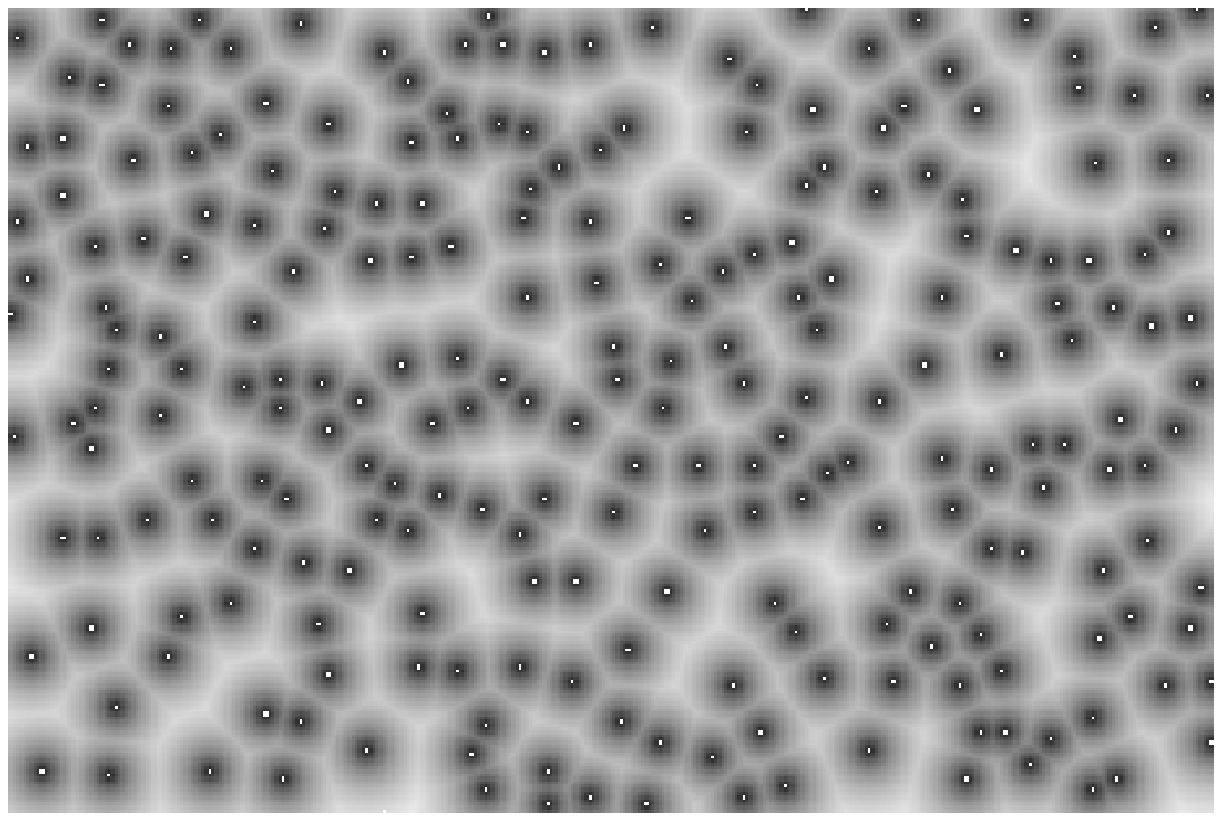} \\
   \vspace{0.2cm}
   (a) \\
   \includegraphics[scale=.55,angle=-90]{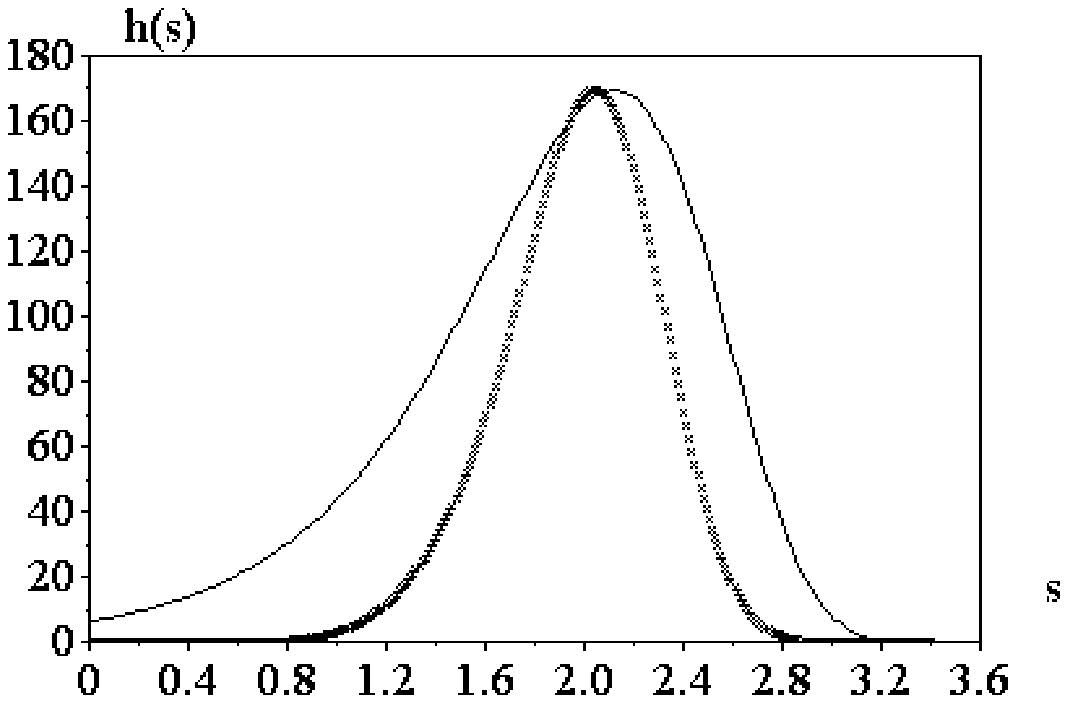} \\
   (b)

   \caption{Spatial distribution of retinal ganglion cells (a) and
   respective $h(s)$ function (b), represented by the dotted line.
   The nearest-neighbor distribution obtained for uniform point
   dispersion with equivalent density is shown by the solid line.
   ~\label{fig:wassle}} \end{center}
\end{figure}

In addition to the above validations performed with synthetic data,
the proposed approach for characterization of multiple-object systems
was also applied to real data regarding the spatial distribution of
the center of mass of the soma of retinal ganglion cells, kindly
provided by H. W\"assle. Reflecting the hexagonal organization of the
photoreceptors characterizing the central portion of the mammals'
retina, the position of these cells at the last retinal layer are also
expected to follow some hexagonal organization, with some degree of
disorder.  The obtained rings around each original point and the
respective function $h(s)$ are shown in Figures~\ref{fig:wassle}(a)
and (b), respectively.  The nearest-neighbor distribution considering
uniformly spaced points for the same density are also shown (solid
line) superposed to the curve $h(s)$ in Figure~\ref{fig:wassle}.  It
is clear from this result that the ganglion cell distribution is
substantially narrower than its respective uniform counterpart,
corroborating a more organized scheme such as the hexagonal spacing.

All in all, a new and effective approach to the characterization of
multiple-object systems, based on artificially-induced percolation,
has been proposed that presents a series of unique and interesting
features.  By considering only the exact distances representable in
discrete spaces, as well as the fact that possibly co-existing point
densities in such systems are typically sampled by using the same
window, it has been possible to derive a powerful normalized tool
capable not only of estimating the density of co-existing object
distributions but also their respective perturbation degree.  In
addition, the number of points present in the observed region is
reflected by the height of the obtained peaks.  It is argued that such
a function provides one of the most informative and accurate means for
quantification of densities and spatial coverage, with the additional
advantage that, unlike alternative schemes based on fractal and
lacunarity, the obtained curve $h(s)$ has an immediate and intuitive
interpretation in terms of point densities.  Although not illustrated
in this work, the proposed method can be immediately applied to
multiple-object systems involving objects with shape and size more
general than points.  In such cases, the typical size of the objects
has also been considered for the correct normalization and
interpretation of the results.  Other perspectives to be pursued
further regard the consideration of the hierarchical structures (a
tree) obtained as a byproduct of the percolation-induced merging
procedure and the possibility of applying the artificially-induced
percolation analysis to characterize weighted networks.  The proposed
approach is currently being applied to characterize the texture of
bones, growing neural systems, and spatial patterns of gene
expression.

\begin{acknowledgments}
The author is grateful to FAPESP (proc. 99/12NPq
(proc. 301422/92-3) and the Human Frontier Science Program for
financial support.
\end{acknowledgments}

\bibliography{geneperc}

\begin{thebibliography}{12}
\expandafter\ifx\csname natexlab\endcsname\relax\def\natexlab#1{#1}\fi
\expandafter\ifx\csname bibnamefont\endcsname\relax
  \def\bibnamefont#1{#1}\fi
\expandafter\ifx\csname bibfnamefont\endcsname\relax
  \def\bibfnamefont#1{#1}\fi
\expandafter\ifx\csname citenamefont\endcsname\relax
  \def\citenamefont#1{#1}\fi
\expandafter\ifx\csname url\endcsname\relax
  \def\url#1{\texttt{#1}}\fi
\expandafter\ifx\csname urlprefix\endcsname\relax\def\urlprefix{URL }\fi
\providecommand{\bibinfo}[2]{#2}
\providecommand{\eprint}[2][]{\url{#2}}

\bibitem[{\citenamefont{Stauffer and Aharony}(1994)}]{Stauffer:94}
\bibinfo{author}{\bibfnamefont{D.}~\bibnamefont{Stauffer}} \bibnamefont{and}
  \bibinfo{author}{\bibfnamefont{A.}~\bibnamefont{Aharony}},
  \emph{\bibinfo{title}{Introduction to Percolation Theory}}
  (\bibinfo{publisher}{Taylor and Francis}, \bibinfo{year}{1994}).

\bibitem[{\citenamefont{A.~J.~Fisher et~al.}(1992)\citenamefont{A.~J.~Fisher,
  Dowrick, and Newman}}]{Binney_etal:94}
\bibinfo{author}{\bibfnamefont{J.~J.~B.} \bibnamefont{A.~J.~Fisher}},
  \bibinfo{author}{\bibfnamefont{N.~J.} \bibnamefont{Dowrick}},
  \bibnamefont{and} \bibinfo{author}{\bibfnamefont{M.~E.~J.}
  \bibnamefont{Newman}}, \emph{\bibinfo{title}{The Theory of Critical
  Phenomena}} (\bibinfo{publisher}{Clarendon Press}, \bibinfo{address}{London},
  \bibinfo{year}{1992}).

\bibitem[{\citenamefont{Albert and Barab\'asi}(2002)}]{Albert_barab:2002}
\bibinfo{author}{\bibfnamefont{R.}~\bibnamefont{Albert}} \bibnamefont{and}
  \bibinfo{author}{\bibfnamefont{A.~L.} \bibnamefont{Barab\'asi}},
  \bibinfo{journal}{Rev. Mod. Phys.} \textbf{\bibinfo{volume}{74}},
  \bibinfo{pages}{47} (\bibinfo{year}{2002}).

\bibitem[{\citenamefont{Ziman}(1979)}]{Ziman:79}
\bibinfo{author}{\bibfnamefont{J.~M.} \bibnamefont{Ziman}},
  \emph{\bibinfo{title}{Models of Disorder}} (\bibinfo{publisher}{Cambridge
  University Press}, \bibinfo{address}{London}, \bibinfo{year}{1979}).

\bibitem[{\citenamefont{Wassle and Riemann}(1978)}]{Wassle:78}
\bibinfo{author}{\bibfnamefont{H.}~\bibnamefont{Wassle}} \bibnamefont{and}
  \bibinfo{author}{\bibfnamefont{H.~J.} \bibnamefont{Riemann}},
  \bibinfo{journal}{Proc. R. Soc. Lond. B. Biol. Sci.}
  \textbf{\bibinfo{volume}{200}}, \bibinfo{pages}{441} (\bibinfo{year}{1978}).

\bibitem[{\citenamefont{Michelsen and de~Raedt}(2001)}]{Raedt01}
\bibinfo{author}{\bibfnamefont{K.}~\bibnamefont{Michelsen}} \bibnamefont{and}
  \bibinfo{author}{\bibfnamefont{H.}~\bibnamefont{de~Raedt}},
  \bibinfo{journal}{Physics Report} \textbf{\bibinfo{volume}{347}},
  \bibinfo{pages}{461} (\bibinfo{year}{2001}).

\bibitem[{\citenamefont{Kaye}(1994)}]{Kaye:94}
\bibinfo{author}{\bibfnamefont{B.~H.} \bibnamefont{Kaye}},
  \emph{\bibinfo{title}{A Random Walk Through Fractal Dimensions}}
  (\bibinfo{publisher}{John Wilhe and Sons}, \bibinfo{year}{1994}).

\bibitem[{\citenamefont{Hovi et~al.}(1996)\citenamefont{Hovi, Aharony,
  Stauffer, and Mandelbrot}}]{Hovi:96}
\bibinfo{author}{\bibfnamefont{J.~P.} \bibnamefont{Hovi}},
  \bibinfo{author}{\bibfnamefont{A.}~\bibnamefont{Aharony}},
  \bibinfo{author}{\bibfnamefont{D.}~\bibnamefont{Stauffer}}, \bibnamefont{and}
  \bibinfo{author}{\bibfnamefont{B.~B.} \bibnamefont{Mandelbrot}},
  \bibinfo{journal}{Phys. Rev. Lett.} \textbf{\bibinfo{volume}{77}},
  \bibinfo{pages}{877} (\bibinfo{year}{1996}).

\bibitem[{\citenamefont{Streicher et~al.}(2000)\citenamefont{Streicher, Donat,
  Strauss, Spoerle, Schughart, and Muller}}]{Streicher:2000}
\bibinfo{author}{\bibfnamefont{J.}~\bibnamefont{Streicher}},
  \bibinfo{author}{\bibfnamefont{M.~A.} \bibnamefont{Donat}},
  \bibinfo{author}{\bibfnamefont{B.}~\bibnamefont{Strauss}},
  \bibinfo{author}{\bibfnamefont{R.}~\bibnamefont{Spoerle}},
  \bibinfo{author}{\bibfnamefont{K.}~\bibnamefont{Schughart}},
  \bibnamefont{and} \bibinfo{author}{\bibfnamefont{G.~B.}
  \bibnamefont{Muller}}, \bibinfo{journal}{Nat. Genet.}
  \textbf{\bibinfo{volume}{25}}, \bibinfo{pages}{147} (\bibinfo{year}{2000}).

\bibitem[{\citenamefont{da~F.~Costa and Jr}(2001)}]{CostaCesar:2001}
\bibinfo{author}{\bibfnamefont{L.}~\bibnamefont{da~F.~Costa}} \bibnamefont{and}
  \bibinfo{author}{\bibfnamefont{R.~M.~C.} \bibnamefont{Jr}},
  \emph{\bibinfo{title}{Shape Analysis and Classification: Theory and
  Practice}} (\bibinfo{publisher}{CRC Press}, \bibinfo{address}{Boca Raton},
  \bibinfo{year}{2001}).

\bibitem[{\citenamefont{Stoyan and Stoyan}(1995)}]{Stoyan:95}
\bibinfo{author}{\bibfnamefont{D.}~\bibnamefont{Stoyan}} \bibnamefont{and}
  \bibinfo{author}{\bibfnamefont{H.}~\bibnamefont{Stoyan}},
  \emph{\bibinfo{title}{Fractals, Random Shapes and Point Fields}}
  (\bibinfo{publisher}{John Wiley and Sons}, \bibinfo{address}{Chichester},
  \bibinfo{year}{1995}).

\bibitem[{\citenamefont{Duda et~al.}(2001)\citenamefont{Duda, Hart, and
  Stork}}]{DudaHart:2001}
\bibinfo{author}{\bibfnamefont{R.~O.} \bibnamefont{Duda}},
  \bibinfo{author}{\bibfnamefont{P.~E.} \bibnamefont{Hart}}, \bibnamefont{and}
  \bibinfo{author}{\bibfnamefont{D.~G.} \bibnamefont{Stork}},
  \emph{\bibinfo{title}{Pattern Classification}}
  (\bibinfo{publisher}{Wiley-Interscience}, \bibinfo{address}{New York},
  \bibinfo{year}{2001}).

\end{thebibliography}

\end{document}